\documentstyle[12pt]{article}
\def\hw{\hbar\omega}
\def\bra#1{\langle#1|}
\def\ket#1{|#1\rangle}

\begin{document}
\sloppy
\sloppy
\sloppy

\begin{flushright}{UT-819\\July, 1998}

\end{flushright}
\vskip 0.5 truecm

\vskip 0.5 truecm

\begin{center}
{\large{\bf  Fluctuation-dissipation theorem  and quantum tunneling with dissipation at finite temperature }}
\end{center}
\vskip .5 truecm
\centerline{\bf Kazuo Fujikawa and Hiroaki Terashima}
\vskip .4 truecm
\centerline {\it Department of Physics,University of Tokyo}
\centerline {\it Bunkyo-ku,Tokyo 113,Japan}
\vskip 0.5 truecm

\begin{abstract}
A reformulation of  the fluctuation-dissipation theorem of Callen and Welton 
is presented in such a manner that the basic idea of Feynman-Vernon and Caldeira -Leggett of using an infinite number of oscillators to simulate the dissipative 
medium is realized manifestly without actually introducing oscillators. If  one 
assumes the existence of a well defined dissipative coefficient $R(\omega)$  
which little depends on the temperature in the energy region we are interested  in , the spontanous and induced emissions as well as  induced 
absorption of these effective oscillators with correct Bose distribution automatically appears.
 Combined with a dispersion relation, we reproduce the 
tunneling formula in the presence of dissipation at finite temperature without referring to an explicit model Lagrangian. The 
fluctuation-dissipation theorem of Callen-Welton is also  generalized to the 
fermionic dissipation (or fluctuation) which allows a transparent 
physical interpretation in terms of second quantized fermionic oscillators. 
This fermionic version of fluctuation-dissipation theorem may become 
relevant  in the analyses of, for example, fermion radiation from a black hole and also  supersymmetry at  the early universe.
\par
\end{abstract}
\ \ \ \ \ \ \ \ ( Physical Review E,  in press)
\section{Introduction}
The fluctuation-dissipation theorem , which relates the spontaneous
fluctuation of "force fileds" in thermal equilibrium to irreversible 
dissipation, provides  a basis  of   statistical mechanics for irreversible
processes which are slightly away from  thermal equilibrium.  The 
fluctuation-dissipation 
theorem has been formulated by various authors in the past [1] - [7]. We
find the formulation by Callen and Welton\cite{callen}  intuitively
understandable and appealing. 
They showed\cite{callen}  that a general form of fluctuation-dissipation
theorem covers a wide range of phenomena such as the Einstein relation for
Brownian motion\cite{einstein}, the Nyquist formula for  voltage
fluctuation in conductors\cite{nyquist}, and the Planck distribution for
photons. In Ref.\cite{fuji}, it was shown that the effect of dissipation on 
quantum tunneling ( or coherence ) 
at {\em zero-temperature} can be formulated on the basis of the fluctuation-dissipation theorem of Callen-Welton and causality and unitarity 
(i.e., dispersion relations) without referring to an explicit form of 
Lagrangian a la Caldeira and Leggett\cite{caldeira}. 

In this  paper, we present a reformulation of the fluctuation-dissipation theorem of Callen and Welton in such a manner that  the basic idea of Feynman-Vernon \cite{feynman} and Caldeira -Leggett\cite{caldeira}, which simulates the dissipative medium by an  infinite number of oscillators, becomes manifest
without actually introducing oscillators. The quantum tunneling at {\em finite 
temperature} is described by this  reformulation.
 Although we use the quantum mechanical Fermi's 
golden rule , the spontanous and induced emissions as well as  induced 
absorption of these effective oscillators with correct Bose distribution automatically appears, if one 
assumes the existence of a well defined dissipative coefficient $R(\omega)$  
which little depends on the temperature in the energy region we are interested  in.  

To be specific, for the simplest case of a hermitian (composite) operator
$Q$, which represents the ( generally complicated ) dynamical freedom of the dissipative medium, we 
have relations
\begin{equation}
 \frac{2}{\pi}\frac{\hw}{2}
    \left[1 + \frac{1}{e^{\beta\hw}- 1}\right]R(\omega)
    =\hbar\int_0^\infty dE\,\rho(E)f(E)\rho(E+\hw)
    \left|\bra{E+\hw}Q\ket{E}\right|^2 
\end{equation}
and 
\begin{equation} 
\frac{2}{\pi}\frac{\hw}{2}
    \left[\frac{1}{e^{\beta\hw}- 1}\right]R(\omega)
    =\hbar\int_0^\infty dE\,\rho(E)f(E)\rho(E-\hw)
    \left|\bra{E-\hw}Q\ket{E}\right|^2
\end{equation}
where $f(E)$ stands for a normalized Boltzmann factor with 
$f(E + \hw)/f(E) = e^{-\beta\hw}$ and $\beta = 1/kT$. 
These formulas are  remarkable: in the right-hand sides of these 
relations , we simply use the Fermi's golden rule for the first order 
perturbation. We recognize the left-hand side of (1) as representing the  
spontaneous and induced 
emissions of second-quantized Bosonic oscillators, and  the left-hand side of (2) as the standard (induced ) absorption formula. Moreover the spectrum of 
the effective 
Bosonic oscillators is characterized by the dissipative coefficient (
resistance) $R(\omega)$. In other words, the presence of $R(\omega)$ necessarily implies the presence of effective oscillators characterized by $R(\omega)$,
as was emphasized in Ref.\cite{fuji}.
 These expressions realize the basic idea of Feynman-Vernon\cite{feynman}
and Caldeira-Leggett\cite{caldeira} without actually introducing an infinite 
number of oscillators , and they naturally satisfiy the  detailed balancing relation. It is shown that these formulas (1) and (2) lead to the conventional fluctuation-dissipation theorem of Callen and Welton. 

Combining (1) and (2) with a dispersion relation, it is shown in the text that we can  reproduce the 
tunneling formula in the presence of dissipation at {\em finite temperature}
 without referring to an explicit model Lagrangian.

We also present a  generalization of the Callen-Welton formula for  fermionic dissipation ( or fluctuation) in contrast to the conventional Bosonic 
dissipation such as in (1) and (2). These formulas for fermionic dissipation, 
though somewhat academic at 
this moment, may become relevant in the analyses of , for example, fermion
emission from a black hole or supersymmetric properties at the early universe.

\section{Reformulation of fluctuation-dissipation theorem} 
\subsection{Microscopic power dissipation}

We first start with a Hamiltonian
\begin{equation}
 H=H_0(Q) + V(Q,q)e^{i\omega t}+V(Q,q)^\dagger e^{-i\omega t},
\label{Hamiltonian}
\end{equation}
where $H_0(Q)$ is the unperturbed Hamiltonian for the dissipative medium,
which has eigenstates
\begin{equation}
  H_0\ket{E_n}=E_n\ket{E_n}.
\end{equation}
The variable $q$  appearing  in $V(Q,q)$  describes the external dynamical freedom which perturbs the dissipative medium. The variable $q$ is treated as 
a classical variable for the moment.

If the dissipative medium   is initially in the state $\ket{E_n}$,
the lowest order transition rate
(transition probability per unit time) is given by the Fermi's golden rule by
treating the last two terms in (3) as small perturbation,
\begin{eqnarray}
  w &=& \frac{2\pi}{\hbar}
 \biggl[ \left|\bra{E_n+\hw}V^\dagger\ket{E_n}\right|^2\,
   \rho(E_n+\hw) \nonumber \\
    & &\qquad\qquad
  +\left|\bra{E_n-\hw}V\ket{E_n}\right|^2\,
   \rho(E_n-\hw) \biggr].
\end{eqnarray}
Since the first term stands for  the absorption of energy $\hw$
and the second term for the emission of energy $\hw$,
the energy absorption rate by the dissipative medium is given  by
\begin{equation}
  2\pi\omega
 \biggl[ \left|\bra{E_n+\hw}V^\dagger\ket{E_n}
   \right|^2\,\rho(E_n+\hw)
  -\left|\bra{E_n-\hw}V\ket{E_n}\right|^2\,
   \rho(E_n-\hw) \biggr].
\end{equation}

If the system is initially in thermal equilibrium at temperature $T$,
we must average over all initial states,
weighting the state $|E_{n}\rangle $ by the (normalized) Boltzmann factor 
$f(E_n)$ which satisfies
\begin{equation}
  \frac{f(E_n+\hw)}{f(E_n)}=e^{-\beta\hw},
   \qquad\qquad \beta=\frac{1}{kT}.
\end{equation}
Then  the energy dissipation per unit time is given by 
\begin{eqnarray}
  P(\omega)&=&2\pi\omega\int_0^\infty dE\,\rho(E)f(E)
 \biggl[ \left|\bra{E+\hw}V^\dagger\ket{E}\right|^2\,
   \rho(E+\hw)  \nonumber \\
   & &\qquad\qquad
  -\left|\bra{E-\hw}V\ket{E}\right|^2\,
   \rho(E-\hw) \biggr].
\end{eqnarray}
where we replaced  the summation over $n$ by an integration over energy.

Now, we assume that $V$ can be written as $V=qQ/2$, namely, the interaction
part in (3) is written as
\begin{equation}
H_{I} = \frac{1}{2}( qe^{i\omega t}Q + Q^{\dagger}\bar{q}e^{-i\omega t})
\end{equation}  
in the spirit of linear response approximation; $q$ is an infinitesimal  complex number  and $Q$ is a bosonic composite 
operator, respectively. Here we allow the operator $Q$ to be non-hermitian 
in general so that we can readily extend our formulation to  fermionic 
dissipation later. Then, we obtain
\begin{eqnarray}
  P(\omega)&=&\frac{\pi\omega}{2}q\bar{q}
   \int_0^\infty dE\,\rho(E)f(E)
 \biggl[ \left|\bra{E+\hw}Q^\dagger\ket{E}\right|^2\,
   \rho(E+\hw)  \nonumber \\
   & &\qquad\qquad
  - \left|\bra{E-\hw}Q\ket{E}\right|^2\,
   \rho(E-\hw) \biggr].
\end{eqnarray}

\subsection{Macroscopic dissipative coefficient $R(\omega)$}

We next define the phenomenological macroscopic dissipative coefficient (resistance) $R(\omega)$ on the basis of the following reasoning\cite{fuji}. We first define  an infinitesimal ( complex ) coordinate in $H_{I}$ (9)
\begin{equation}
q(t) = qe^{i\omega t}
\end{equation}
The existence of the energy dissipation into the medium (10) induced by the 
external perturbation $q(t)$ 
suggests the presence of a dissipative  force (reaction) acting on the variable $Re q(t)$, which  oscillates with frequency $\omega$, 
\begin{equation}
F = - R(\omega) Re \dot{q}(t)
\end{equation}
where  $R(\omega)$ is a real function.  Note that $ H_{I}= Q Re q(t)$ in (9)
for a hermitian operator $Q$ ; this shows that $Re q(t)$ is a natural classical
counter part of the ( hermitian ) quantum variable $\hat{q}$ to describe the 
macroscopic quantum system $H_{0}(q)$.   

The power dissipation per unit time generated by this phenomenological 
reactive  force is given by 
\begin{eqnarray}
P(\omega) &=& - \overline{Re F Re \dot{q}(t)}\nonumber\\
      &=& R(\omega) \overline{(Re \dot{q}(t))^{2}}\nonumber\\
      &=& \frac{1}{2}R(\omega) \overline {(Re \dot{q}(t))^{2} + (Im \dot{q}(t))^{2}}\nonumber\\
      &=&  \frac{\omega^{2}}{2}R(\omega) q\bar{q}
\end{eqnarray}
where the overline indicates time averaging.

Combining (10) and (13), we obtain the microscopic expression for the dissipative coefficient ( resistance ) $R(\omega)$
\begin{eqnarray}
  R(\omega)&=&\frac{\pi}{\omega}
   \int_0^\infty dE\,\rho(E)f(E)
 \biggl[ \left|\bra{E+\hw}Q^\dagger\ket{E}\right|^2\,
   \rho(E+\hw)  \nonumber \\
   & &\qquad\qquad \nonumber
  - \left|\bra{E-\hw}Q\ket{E}\right|^2\,
   \rho(E-\hw) \biggr] \label{R} \\
   &=& \frac{\pi}{\omega}(1 - e^{-\beta\hw})
   \int_0^\infty dE\,\rho(E)f(E)\rho(E+\hw)
    \left|\bra{E+\hw}Q^\dagger\ket{E}\right|^2.\nonumber\\
\end{eqnarray}
From this expression of $R(\omega)$, we  find the basic relations,
\begin{eqnarray}
 \frac{2}{\pi}\frac{\hw}{2}
    \left[1 + \frac{1}{e^{\beta\hw} - 1}\right]R(\omega)
    &=&\hbar\int_0^\infty dE\,\rho(E)f(E)\rho(E+\hw)
    \left|\bra{E+\hw}Q^\dagger\ket{E}\right|^2,\nonumber\\ 
      \label{Qd1} \\
 \frac{2}{\pi}\frac{\hw}{2}
    \left[\frac{1}{e^{\beta\hw} - 1}\right]R(\omega)
    &=&\hbar\int_0^\infty dE\,\rho(E)f(E)\rho(E-\hw)
    \left|\bra{E-\hw}Q\ket{E}\right|^2.\nonumber\\
      \label{Q1}
\end{eqnarray}
Eq.(\ref{Qd1}) stands for the absorption of $\hbar\omega$ by $Q^\dagger$
and Eq.(\ref{Q1}) stands for the  emission of $\hbar\omega$ by $Q$.

These relations (\ref{Qd1}) and (\ref{Q1}) stand for the proto-type of the 
fluctuation-dissipation theorem of Callen and Welton. The fluctuation-dissipation theorem as it stands is a mathematical identity and contains no physical
contents by itself. What is remarkable is  that we obtain  highly non-trivial
relations in (\ref{Qd1}) and (\ref{Q1}) if one {\em assumes} that the 
dissipative coefficient $R(\omega)$ little depends on the temperature in the 
region we are interested in. We can recognize the left-hand side of (\ref{Qd1}) as standing for the spontaneous and induced emissions of the second quantized  bosonic oscillator with frequency $\omega$ into the dissipative medium,
whereas the left-hand side of (\ref{Q1}) is recognized as the (induced) absorption  of these oscillators from the dissipative medium at temperature $T$. Moreover, the spectrum 
of these effective oscillators is precisely specified by the dissipative
coefficient $R(\omega)$. In particular, there is  no effective oscillators of 
$\omega$ for which $R(\omega)$ vanishes. These properties realize the basic
physical idea of Feynman-Vernon\cite{feynman} and Caldeira-Leggett\cite{caldeira}, which simulates the 
dissipative medium by an infinite number of oscillators. In our approach
this physical idea is realized by a simple application of the Fermi's golden
rule combined with the temperature independence of the dissipative coefficient
$R(\omega)$ without actually introducing oscillators. We emphasize that these oscillators are effective ones and there
do not exist such real oscillators inside the dissipative medium in 
general. It is shown in next section that the quantum tunneling with 
dissipation at {\em finite temperature} is formulated on the basis of      
(\ref{Qd1}) and (\ref{Q1}) and dispersion relations (i.e., unitarity and 
causality) without referrring to the explicit model Lagrangian of Caldeira
and Leggett.

It is obvious from Eq.(\ref{Hamiltonian})
that the change $\omega\to-\omega$ corresponds
to the exchange $V\leftrightarrow V^\dagger$.
By definition of $R(\omega)$ in (\ref{R}), we then obtain the expression
\begin{eqnarray}
  R(-\omega)&=& \frac{\pi}{\omega}
   \int_0^\infty dE\,\rho(E)f(E)
 \biggl[ \left|\bra{E+\hw}Q\ket{E}\right|^2\,
   \rho(E+\hw)  \nonumber \\
   & &\qquad\qquad
  - \left|\bra{E-\hw}Q^\dagger\ket{E}\right|^2\,
   \rho(E-\hw) \biggr] \\
   &=&\frac{\pi}{\omega}(1 -  e^{-\beta\hw})
   \int_0^\infty dE\,\rho(E)f(E)\rho(E+\hw)
    \left|\bra{E+\hw}Q\ket{E}\right|^2.
\end{eqnarray}
Clearly, $R(-\omega)= R(\omega)$ in the case of
a hermitian operator, $Q=Q^{\dagger}$. 

For $R(-\omega)$ with $\omega > 0$, we obtain relations similar to (15) and 
(16),
\begin{eqnarray}
 \frac{2}{\pi}\frac{\hw}{2}
    \left[1 + \frac{1}{e^{\beta\hw} - 1}\right]R(-\omega)
    &=&\hbar\int_0^\infty dE\,\rho(E)f(E)\rho(E+\hw)
    \left|\bra{E+\hw}Q\ket{E}\right|^2,\nonumber\\ 
      \label{Q2} \\
 \frac{2}{\pi}\frac{\hw}{2}
    \left[\frac{1}{e^{\beta\hw} - 1}\right]R(-\omega)
    &=&\hbar\int_0^\infty dE\,\rho(E)f(E)\rho(E-\hw)
    \left|\bra{E-\hw}Q^\dagger\ket{E}\right|^2.\nonumber\\
     \label{Qd2}
\end{eqnarray}
Eq.(\ref{Q2}) stands for the  absorption of $\hbar \omega$ by $Q$
and Eq.(\ref{Qd2}) stands for the  emission of $\hbar \omega$ by $Q^\dagger$
from the dissipative medium, respectively.

\subsection{Fluctuation-dissipation theorem}

Finally, we formulate  the fluctuation-dissipation theorem of Callen and 
Welton.
From Eqs.(\ref{Qd1}) and (\ref{Qd2}), we have after integration over $\omega$
( and changing the order of integration over $E$ and $\omega$)
\begin{eqnarray}
 \langle QQ^\dagger\rangle 
 &\equiv& \int ^{\infty}_{0}\langle E| QQ^{\dagger}|E\rangle\rho (E) f(E)dE
  \nonumber\\
  &=&\int ^{\infty}_{0}dE\rho (E)f(E)\{\int ^{\infty}_{0}|\langle E+\hbar
     \omega|Q^{\dagger}|E\rangle |^{2}\rho (E+\hbar\omega)d(\hbar\omega)
     \nonumber\\
  && +\int ^{\infty}_{0}|\langle E-\hbar
     \omega|Q^{\dagger}|E\rangle |^{2}\rho (E-\hbar\omega)d(\hbar\omega)\} 
     \nonumber\\    
  &=&\frac{2}{\pi}\int_0^\infty d\omega \frac{\hw}{2}
    \left\{ \left[1 + \frac{1}{e^{\beta\hw} - 1}\right]R(\omega)
    + \left[\frac{1}{e^{\beta\hw}- 1}\right]R(-\omega) \right\}\nonumber\\
\label{FD1}
\end{eqnarray}
and similarly from Eqs.(\ref{Q1}) and (\ref{Q2}),
\begin{equation}
  \langle Q^\dagger Q\rangle
    =\frac{2}{\pi}\int_0^\infty d\omega \frac{\hw}{2}
    \left\{ \left[\frac{1}{e^{\beta\hw} - 1}\right]R(\omega)
    + \left[1 + \frac{1}{e^{\beta\hw}- 1}\right] R(-\omega) \right\}
\end{equation}
Furthermore, 
\begin{equation}
 \frac{1}{2} \langle Q^\dagger Q + QQ^\dagger\rangle
    =\frac{2}{\pi}\int_0^\infty d\omega\;
    E(\omega,T)\; \left[\frac{R(\omega)+R(-\omega)}{2}\right],
\label{FD3}
\end{equation}
where
\begin{equation}
    E(\omega,T)=\frac{\hw}{2}+\frac{\hw}{e^{\beta\hw} - 1}.
\end{equation}
If Q is a hermitian operator,
$Q=Q^\dagger$, we have $R(-\omega)=R(\omega)$, and 
Eqs.(\ref{FD1})--(\ref{FD3}) all reduce to the familiar form of 
the fluctuation-dissipation theorem of Callen and Welton\cite{callen},
\begin{equation}
 \langle Q^2\rangle
    =\frac{2}{\pi}\int_0^\infty d\omega\;
    E(\omega,T)\; R(\omega).
\end{equation}
The fluctuation $ \langle Q^2\rangle$ in the left-hand side is expressed in
terms of the dissipative coefficient $R(\omega)$ with a universal kernel
$E(\omega,T)$. The fluctuation-dissipation theorem  is regarded as a sum rule 
relating two quantities $ \langle Q^2\rangle$ and $R(\omega)$, which are measured independently. 

In Ref.\cite{fuji}, we used the final formula (25) and thus only the {\em 
zero temperature} case of  quantum tunneling with dissipation was formulated  
in a model independent manner. The relations (\ref{Qd1}) and (\ref{Q1})
apparently contain more information, and they  allow us to formulate the tunneling with dissipation at {\em finite} temperature.

\section{Quantum tunneling with dissipation at finite temperature}

We start with the total Hamiltonian
\begin{equation}
H = H_{0}(Q) + H_{0}(q) + H_{I}(q,Q)
\end{equation}
where $H_{0}(q)$ describes the unperturbed Hamiltonian of the quantum system
we are interested in
\begin{equation}
H_{0}(q) = \frac{1}{2M}p^{2} + V(q)
\end{equation}
 and $H_{I}(q,Q)$ stands for the interaction Hamiltonian in (10) 
\begin{equation}
H_{I}(q,Q) = qQ
\end{equation} 
but now the variable $q$ is promoted to a hermitian quantum operator; the 
explicit time dependence disappears in the Schroedinger picture. 
We also choose a hermitian $Q$, $Q^{\dagger} = Q$,  in conformity with the convention of the standard Caldeira-Leggett model\cite{caldeira}. In the context 
of macroscopic  
quantum tunneling (or to be more precise, quantum coherence),  we choose
$V(q)$ in (28) as standing for a symmetric double well potential.
$H_{0}(Q)$ describes the dissipative medium as before, but we do not need an 
explicit form of $H_{0}(Q)$ and   the dynamical properties of  $H_{0}(Q)$ are
 indirectly specified by our fluctuation-dissipation theorem (15) and (16).

We now start with the eigenstates of $H_{0}(q)$
\begin{equation}
H_{0}(q)|n\rangle = E_{n}|n\rangle
\end{equation}
and consider the transition probability for $n \rightarrow m + \hbar\omega$
by emitting energy $\hbar\omega$ to the dissipative medium, which is assumed 
to be in thermal equilibrium with temperature $T$.  
The transition probability for this process is given by the lowest order 
perturbation of $H_{I}(q,Q)$ as 
\begin{eqnarray}
&&w(n \rightarrow m + \hbar\omega)\nonumber\\
&&= \frac{2\pi}{\hbar}|\langle m|q|n\rangle|^{2}\int_{0}^{\hbar\Lambda}[
 \int_{0}^{\infty}\rho ( E + \hbar\omega )|\langle E+\hbar\omega|Q|E\rangle|^{2} \rho(E)f(E)dE]\nonumber\\
&&\ \ \ \times \delta (E_{n} - E_{m} - \hbar\omega)d(\hbar\omega)\nonumber\\
&&=\frac{2\pi}{\hbar}|\langle m|q|n\rangle|^{2}\frac{2}{\pi}\int_{0}^{\hbar\Lambda}\frac{\hbar\omega}{2}[ 1 + \frac{1}{e^{\hbar\omega/kT}-1}]\frac{R(\omega)}{\hbar}\delta (E_{n} - E_{m} - \hbar\omega)d(\hbar\omega)\nonumber\\
\end{eqnarray}
where we used eq.(15) for a hermitian $Q$. We also introduced an explicit 
cut-off $\hbar\Lambda$ of  effective frequency of the dissipative medium,
which could be included in the definition of $R(\omega)$.

At finite temperature we also have an absorption probability 
\begin{eqnarray}
&&w(n + \hbar\omega \rightarrow m)\nonumber\\
&&= \frac{2\pi}{\hbar}|\langle m|q|n\rangle|^{2}\int_{0}^{\hbar\Lambda}[
 \int_{0}^{\infty}\rho ( E - \hbar\omega )|\langle E-\hbar\omega|Q|E\rangle|^{2} \rho(E)f(E)dE]\nonumber\\
&&\ \ \ \times \delta (E_{n} - E_{m} + \hbar\omega)d(\hbar\omega)\nonumber\\
&&=\frac{2\pi}{\hbar}|\langle m|q|n\rangle|^{2}\frac{2}{\pi}\int_{0}^{\hbar\Lambda}\frac{\hbar\omega}{2}[ \frac{1}{e^{\hbar\omega/kT}-1}]\frac{R(\omega)}{\hbar}\delta (E_{n} - E_{m} + \hbar\omega)d(\hbar\omega)\nonumber\\
\end{eqnarray}
where we used (16). It is interesting that these formulas satisfy the detailed
balancing relation
\begin{equation}
e^{\hbar\omega/kT}w(n \rightarrow m + \hbar\omega)= w(m + \hbar\omega \rightarrow n)
\end{equation}
with $\hbar\omega = E_{n} - E_{m}$. 

We next  define the half-width of the state $|n\rangle$ for emission( when we prepare the state $|n\rangle$ at $t=0$)
\begin{equation}
\frac{1}{2}\Gamma_{n}^{(+)} = \frac{1}{2}\hbar \sum_{m}w(n \rightarrow m
+ \hbar\omega)
\end{equation}
and the corresponding one for absorption
\begin{equation}
\frac{1}{2}\Gamma_{n}^{(-)} = \frac{1}{2}\hbar \sum_{m}w(n + \hbar\omega 
\rightarrow m )
\end{equation}
It can be confirmed that eq. (33) gives $(1/2)\Gamma_{n}= (1/2)\hbar\eta/M$
 at $T=0$ for a simple harmonic oscillator $H_{0}(q) = (1/2M)p^{2} + 
(M\omega^{2}/2)q^{2}$ and Ohmic dissipation $R(\omega) = \eta $\cite{fuji}; 
this expression of $\Gamma_{n}$ is consistent with a damped oscillator 
$M\ddot{q} + \eta\dot{q} + M\omega^{2}q = 0$, which in turn justifies the 
normalization of $H_{I}$ in (28).  

The basic idea in  our attempt to reproduce the results of the Caldeira-
Leggett model without introducing an auxiliary infinite number of oscillators
is to write dispersion relations, which relate the imaginary part of energy
eigenvalue to the corresponding real part. The imaginary parts are evaluated 
by means of the fluctuation-dissipation theorem as in (33) and (34). 

We thus write a generalization of the dispersion relation for the self-energy
correction $\Sigma_{n}(E)$ to the energy eigenvalue $E_{n}$ as
\begin{eqnarray}
\Sigma_{n}(E) &=& \frac{1}{\pi}\int_{0}^{\infty}\frac{Im \Sigma_{n}^{(+)}(E^{\prime})dE^{\prime}}{E^{\prime} - E - i\epsilon}\nonumber\\
&+& \frac{1}{\pi}\int_{-\hbar\Lambda}^{\infty}\frac{Im \Sigma_{n}^{(-)}(E^{\prime})dE^{\prime}}{E^{\prime} - E - i\epsilon}   
\end{eqnarray}
where
\begin{eqnarray}
Im \Sigma_{n}^{(+)}(E)&\equiv& \frac{1}{2}\Gamma_{n}^{(+)}(E)\nonumber\\
      &=& \sum_{m} |\langle m|q|n\rangle|^{2}\int_{0}^{\hbar\Lambda}\hbar\omega [1 +  \frac{1}{e^{\hbar\omega/kT}-1}]\frac{R(\omega)}{\hbar}\delta (E - E_{m} - \hbar\omega)d(\hbar\omega)\nonumber\\
\end{eqnarray}
and 
\begin{eqnarray}
Im \Sigma_{n}^{(-)}(E)&\equiv& \frac{1}{2}\Gamma_{n}^{(-)}(E)\nonumber\\
 &=& \sum_{m} |\langle m|q|n\rangle|^{2}\int_{0}^{\hbar\Lambda}\hbar\omega [ \frac{1}{e^{\hbar\omega/kT}-1}]\frac{R(\omega)}{\hbar}\delta (E - E_{m} + \hbar\omega)d(\hbar\omega)\nonumber\\
\end{eqnarray}
Note that the lower bound of the integration range in the second term in 
eq.(35) starts at  $-\hbar\Lambda$ due to the definition in (37).  We thus 
obtain
\begin{eqnarray}
\Sigma_{n}(E)&=& \sum_{m}|\langle m|q|n\rangle|^{2}\frac{1}{\pi}\int_{0}^{\hbar
\Lambda}\{ \frac{\hbar\omega}{E_{m} + \hbar\omega - E -i\epsilon}
[1 +  \frac{1}{e^{\hbar\omega/kT}-1}] \nonumber\\
&+& \frac{\hbar\omega}{E_{m} - \hbar\omega - E -i\epsilon}
[\frac{1}{e^{\hbar\omega/kT}-1}]\}
\frac{R(\omega)}{\hbar}d(\hbar\omega)
\end{eqnarray}
\\
{\bf Ohmic dissipation}\\

For the Ohmic dissipation, $R(\omega)\equiv \eta = constant$, in which we specialize from now on, we have the real part of the energy shift from (38) as 
\begin{eqnarray}
&&Re \Sigma_{n}(E_{n})\nonumber\\
&&=\frac{\eta}{\hbar\pi}\sum_{m}|\langle m|q|n\rangle|^{2}\hbar\Lambda
   \nonumber\\
&&-\frac{\eta}{\hbar\pi}\sum_{m}|\langle n|q|m\rangle|^{2}(E_{m} - E_{n})
\int_{0}^{\hbar\Lambda}P\{\frac{1}{\hbar\omega + E_{m} - E_{n}} \nonumber\\
&&+ [\frac{1}{\hbar\omega + E_{m} - E_{n}} + \frac{1}{\hbar\omega - E_{m} + E_{n}}]\frac{1}{e^{\hbar\omega/kT}-1}\}d(\hbar\omega)
\end{eqnarray}
where $P$ stands for the principal value prescription. 
After subtracting the first term proportional to $\hbar\Lambda$ as a renormalization of the potential\cite{fuji2} following the prescription of Caldeira and 
Leggett\cite{caldeira}, the real part $Re \Sigma_{n}(E_{n})$ is rewritten as 
\begin{eqnarray}
&&Re \Sigma_{n}(E_{n}) = - \frac{\eta}{\hbar\pi}\sum_{m}|\langle n|q|m\rangle|^{2}(E_{m} - E_{n})\nonumber\\
&&\times \int_{0}^{\hbar\Lambda}\frac{1}{2}\{ [\frac{1}{\hbar\omega + E_{m} - E_{n}} + \frac{1}{\hbar\omega - E_{m} + E_{n}}](1 + \frac{2}{e^{\hbar\omega/kT}
- 1})
\}d(\hbar\omega)\nonumber\\
&&=- \frac{\eta}{\hbar\pi}\sum_{m}|\langle n|q|m\rangle|^{2}(E_{m} - E_{n})
\nonumber\\ 
&&\times \int_{0}^{\hbar\Lambda}\frac{1}{2}P [\frac{1}{\hbar\omega + E_{m} - E_{n}} + \frac{1}{\hbar\omega - E_{m} + E_{n}}]\coth(\frac{\beta\hbar\omega)}{2})
d(\hbar\omega) 
\end{eqnarray}
which agrees with the result of the field theoretical formulation of the 
Caldeira-Leggett model\cite{fuji2}. It should be noted that the vacuum
fluctuation term ( spontaneous emission ) in (1) plays a central role
in our application, unlike the conventional applications of the fluctuation-
dissipation theorem where the vacuum fluctuation is usually subtracted \cite{callen}.

For $\beta = 1/kT \rightarrow \infty $, one naturally recovers the zero
temperature result\cite{fuji}. For the {\em two-level approximation}, which is valid for 
the lowest two levels in a deep double well potential, we have the result\cite{fuji2}
\begin{equation}
Re \Sigma_{2}(E_{2}) - Re \Sigma_{1}(E_{1}) \simeq
  \epsilon \bar{\eta}\ln (e^{-2}\beta\hbar\Lambda)
\end{equation}
with the zeroth order energy difference $\epsilon \equiv E_{2} - E_{1}$ and 
$\bar{\eta} = \frac{2\eta}{\pi\hbar}|\langle 0|q|1\rangle|^{2}$ for the 
temperature $\epsilon \ll 1/\beta \ll \hbar\Lambda$. One can confirm the absence of $\epsilon \ln \epsilon$ dependence by splitting the integration range in (40)
into $[0,\hbar\Lambda] = [0,a] + [a,\hbar\Lambda ]$ with $\epsilon \ll a \ll 1/\beta \ll \hbar\Lambda$. The energy splitting ( {\em order parameter} of quantum coherence)  corrected by the dissipation  is then given by
\begin{eqnarray}
\epsilon_{(1)} &=& (E_{2} - Re \Sigma_{2}(E_{2})) - (E_{1} - Re \Sigma_{1}(E_{1}))\nonumber\\
& \simeq& \epsilon [ 1 - \bar{\eta}\ln (e^{-2}\beta\hbar\Lambda)]
 \nonumber\\
&\simeq& \epsilon [(e^{-2})\beta\hbar\Lambda]^{-\bar{\eta}}  
\end{eqnarray}
after the renormalization group improvement. This result, which suggests
the suppression of quantum coherence $\epsilon_{(1)} \ll \epsilon$ for
$\beta \hbar\Lambda \gg 1$,  is in agreement with the dilute instanton 
analysis for the case of Ohmic dissipation\cite{bray-moore}. Eq.(42),
when compared with the result $\epsilon_{(1)} = \epsilon [\hbar\omega_{0}/\epsilon]^{-\bar{\eta}}$ or $\epsilon_{(1)} = \epsilon [\hbar\omega_{0}/\epsilon]^{-\bar{\eta}/(1 - \bar{\eta})}$ at $T=0$\cite{bray-moore}\cite{fuji2}, shows that the infrared cut-off , which was originally provided by $\epsilon$, is 
replaced by $1/\beta$. We can thus analyze the quantum coherence without 
referring to an explicit model Lagrangian.   

Although the mathematical basis of the dispersion relation at  finite 
temperature is less solid compared with the one  at  zero
temperature, our relation (35) is justisfied  in the present linear response 
approximation in a limited temperature region since it coincides with the second order perturbation theory  combined with (15) and (16)(or (30) and (31));
we note that the diagonal matrix element $\langle n|q|n\rangle = 0$ for a 
specific double-well potential.  

\section{Fluctuation-dissipation theorem for fermionic dissipation}
We now discuss a fermionic version of the fluctuation-dissipation theorem of 
Callen and Welton starting with eq.(9). The operators $Q$ and $Q^{\dagger}$
are now taken to be fermionic operators and $q$ and $\bar{q}$ are Grassmann numbers which 
satisfy
\begin{equation}
q\bar{q} = - \bar{q}q,\ \ \  q^{2} = 0,\ \ \  \bar{q}^{2} = 0
\end{equation}
and 
\begin{equation}
Qq = - qQ,\ \ \  Q\bar{q} = - \bar{q}Q,\ \ \  Q^{2} = 0,\ \ \  (Q^{\dagger})^{2}=0
\end{equation}
Because of $Q^{2} = 0$, the real fermionic case is trivial, and we consider a
complex $Q$ which satisfy $QQ^{\dagger}\neq 0$. In a relativistic notation of 
4-dimensional space-time, our $Q$ is regarded as one of the components  of 
(composite) two-component complex spinor $Q_{\alpha}, \alpha = 1, 2$. (Instead,
one may also consider a set of real fermionic operators, $Q_{1}$ and $Q_{2}$
with $Q_{1}^{\dagger}=Q_{1},Q_{2}^{\dagger}=Q_{2}$  corresponding to a Majorana
spinor, which satisfy $Q_{1}Q_{2}\neq 0$).    

Eq.(10) is now replaced by
\begin{eqnarray}
  P(\omega)&=&\frac{\pi\omega}{2}q\bar{q}
   \int_0^\infty dE\,\rho(E)f(E)
 \biggl[ \left|\bra{E+\hw}Q^\dagger\ket{E}\right|^2\,
   \rho(E+\hw)  \nonumber \\
   & &\qquad\qquad
  + \left|\bra{E-\hw}Q\ket{E}\right|^2\,
   \rho(E-\hw) \biggr].
\end{eqnarray}
Note the relative sign of two terms in (45), which arises from the Grassmann
nature of $q$ and $\bar{q}$.
We adopt the definition of the dissipative coefficient $P(\omega) = \frac{\omega^{2}}{2}R_{f}(\omega) q\bar{q} $ as in eq.(13). We thus  obtain
\begin{eqnarray}
R_{f}(\omega)&=&\frac{\pi}{\omega}
   \int_0^\infty dE\,\rho(E)f(E)
 \biggl[ \left|\bra{E+\hw}Q^\dagger\ket{E}\right|^2\,
   \rho(E+\hw)  \nonumber \\
   & &\qquad\qquad \nonumber
  + \left|\bra{E-\hw}Q\ket{E}\right|^2\,
   \rho(E-\hw) \biggr] \label{R} \\
   &=& \frac{\pi}{\omega}(1 +  e^{-\beta\hw})
   \int_0^\infty dE\,\rho(E)f(E)\rho(E+\hw)
    \left|\bra{E+\hw}Q^\dagger\ket{E}\right|^2.  
\end{eqnarray}
From this expression of $R_{f}(\omega)$, we  find the basic relations
\begin{eqnarray}
 \frac{2}{\pi}\frac{\hw}{2}
    \left[1 - \frac{1}{e^{\beta\hw} + 1}\right]R_{f}(\omega)
    &=&\hbar\int_0^\infty dE\,\rho(E)f(E)\rho(E+\hw)
    \left|\bra{E+\hw}Q^\dagger\ket{E}\right|^2,\nonumber \\
\\
 \frac{2}{\pi}\frac{\hw}{2}
    \left[\frac{1}{e^{\beta\hw} + 1}\right]R_{f}(\omega)
    &=&\hbar\int_0^\infty dE\,\rho(E)f(E)\rho(E-\hw)
    \left|\bra{E-\hw}Q\ket{E}\right|^2.\nonumber\\
\end{eqnarray}
The left-hand side of (47) is regarded as the spontaneous and induced
emissions of (effective) fermionic oscillators, and the ralative minus sign accounts 
for the Fermi statistics. The left-hand side of (48) is regarded as the 
(induced) absorption of the fermionic oscillators. 

Eq.(17) is now replaced by
\begin{equation}
 \bar{R}_{f}(-\omega)=  \frac{\pi}{\omega}(1 +  e^{-\beta\hw})
   \int_0^\infty dE\,\rho(E)f(E)\rho(E+\hw)
    \left|\bra{E+\hw}Q\ket{E}\right|^2.
\end{equation}
where we defined $P(-\omega) = \frac{\omega^{2}}{2}\bar{R}_{f}(\omega)\bar{q}
q $ by changing the order of $q$ and $\bar{q}$.
For $\bar{R}_{f}(-\omega)$ with $\omega > 0$, we thus obtain the  relations 
\begin{eqnarray}
 \frac{2}{\pi}\frac{\hw}{2}
    \left[1 - \frac{1}{e^{\beta\hw} + 1}\right]\bar{R}_{f}(-\omega)
    &=&\hbar\int_0^\infty dE\,\rho(E)f(E)\rho(E+\hw)
    \left|\bra{E+\hw}Q\ket{E}\right|^2, \nonumber\\
\\
 \frac{2}{\pi}\frac{\hw}{2}
    \left[\frac{1}{e^{\beta\hw} + 1}\right]\bar{R}_{f}(-\omega)
    &=&\hbar\int_0^\infty dE\,\rho(E)f(E)\rho(E-\hw)
    \left|\bra{E-\hw}Q^\dagger\ket{E}\right|^2.\nonumber\\
\end{eqnarray} 

Combining these relations we finally obtain the fluctuation-dissipation
theorem for fermionic dissipation ( or fluctuation) as
\begin{eqnarray}
 \langle QQ^\dagger\rangle 
 &\equiv& \int ^{\infty}_{0}\langle E| QQ^{\dagger}|E\rangle\rho (E) f(E)dE
  \nonumber\\
  &=&\int ^{\infty}_{0}dE\rho (E)f(E)\{\int ^{\infty}_{0}|\langle E+\hbar
     \omega|Q^{\dagger}|E\rangle |^{2}\rho (E+\hbar\omega)d(\hbar\omega)
     \nonumber\\
  && +\int ^{\infty}_{0}|\langle E-\hbar
     \omega|Q^{\dagger}|E\rangle |^{2}\rho (E-\hbar\omega)d(\hbar\omega)\} 
     \nonumber\\    
  &=&\frac{2}{\pi}\int_0^\infty d\omega \frac{\hw}{2}
    \left\{ \left[1 - \frac{1}{e^{\beta\hw} + 1}\right]R_{f}(\omega)
    + \left[\frac{1}{e^{\beta\hw}+ 1}\right]\bar{R}_{f}(-\omega) \right\}\nonumber\\
\end{eqnarray}
and similarly 
\begin{equation}
  \langle Q^\dagger Q\rangle
    =\frac{2}{\pi}\int_0^\infty d\omega \frac{\hw}{2}
    \left\{ \left[\frac{1}{e^{\beta\hw} + 1}\right]R_{f}(\omega)
    + \left[1 - \frac{1}{e^{\beta\hw}+ 1}\right] \bar{R}_{f}(-\omega) \right\}
\end{equation}
Furthermore, 
\begin{equation}
 \frac{1}{2} \langle Q^\dagger Q - QQ^\dagger\rangle
    =\frac{2}{\pi}\int_0^\infty d\omega\;
    E_{f}(\omega,T)\; \left[\frac{R_{f}(\omega) - \bar{R}_{f}(-\omega)}{2}\right],
\end{equation}
where
\begin{equation}
    E_{f}(\omega,T)=-\frac{\hw}{2}+\frac{\hw}{e^{\beta\hw} + 1}.
\end{equation}
Our sign convention of the dissipative coefficients $R_{f}(\omega)$
and $\bar{R}_{f}(-\omega)$ is chosen so that $R_{f}(\omega)\geq 0$ and 
$\bar{R}_{f}(-\omega)\geq 0$.

In practice, the composite operator $Q$ may carry a well-defined fermion
number and the fermion number may be conserved. Moreover the production of
an anti-particle ( or hole state ) may be suppressed; in such a case, one may 
set  $\bar{R}_{f}(-\omega)= 0$ in our formulas. The physical content of the 
fluctuation-dissipation theorem for fermionic dissipation, as is formulated 
here, is that the 
thermal average of the (composite) operator $QQ^{\dagger}$ or $Q^{\dagger}Q$
, which characterizes the fluctuation,  is represented in terms of effective 
fermionic excitations with their 
spectrum being specified by $R_{f}(\omega)$; the parameter $R_{f}(\omega)$
in turn  characterizes the energy dissipation  into the dissipative medium. 
In this context, the presence of energy dissipation with $R_{f}(\omega)$  
necessarily leads to  the 
presence of effective (or collective) fermionic excitations; this property is
analogous to the Nambu-Goldstone theorem for spontaneous symmetry breakdown,
which asserts an inevitable appearance of massless excitations when 
continuous symmetry is spontaneously broken, as was emphasized in Ref.\cite{fuji}. The difference is that these collective excitations in the present context
are {\em effective} ones.

We finally  comment on a technical complication in deriving (46).   
The  calculation (46) may appear  straightforward and identical to (14).
However, a closer examination reveals that an additional assumption is in
fact involved: 
If $\ket{E}$ is a bosonic state (with even ``G parity''\cite{gso}
), {\em i.e.},
\begin{equation}
  q\;\ket{E}=\ket{E}\;q
\end{equation}
then $Q\ket{E}$ is a fermionic state (odd G parity), {\em i.e.},
\begin{equation}
  q\;\left(Q\ket{E}\right)=-\left(Q\ket{E}\right)\;q
\end{equation}
and vice versa.
Since $\bra{E\pm\hw}Q\ket{E}$ is not a Grassmann variable
but rather an  ordinary number,
$\ket{E\pm\hw}$ should have different G parity from $\ket{E}$.
That is, the interaction $qe^{i\omega t}Q+Q^\dagger\bar{q}e^{-i\omega t}$
would induce not only the energy shift $E\to E\pm\hw$
but also the change of G parity.
This fact prevents one from obtaining  Eq.(46) in a naive way,
because the state $\ket{E}$ obtained  from $\ket{E-\hw}$ by a shift in $E$,
$E \rightarrow E+\hbar \omega$, has a different G
parity from the original $\ket{E}$.
In obtaining (46) we assume that 
the initial state $\ket{E}$ has even G parity
with probability $1/2$ and
odd G parity with probability $1/2$. This assumption is consistent with the
notion of dissipation, which implies that the "radiation" carries away or
injects 
a small amount of energy specified by $\hbar\omega$ at a time. 
We can thus obtain Eq.(46) by averaging over G parity also.

\section{Conclusion}
The main purpose of the present paper is to point out the remarkable relations (1) and (2), which arise from a simple application of the Fermi's golden rule
and the weak temperature dependence of the dissipative coefficient, and their 
applications to quantum tunneling with dissipation. In the conventional 
treatment of the fluctuation-dissipation theorem in (25),   
the vacuum fluctuation is often subtracted away by simply saying that
it is not observable\cite{takagi}. In contrast, the term corresponding to the 
vacuum fluctuation plays a central role in our application, since it describes 
the spontaneous emission of effective excitations into the dissipative medium;
it is  thus the only effect remaining at the vanishing temperature. 

Our dispersion relation for the self-energy (35), which is essentially equivalent to the second order perturbation theory in linear response approximation,
then gives rise to a change in the real part of the tunneling energy splitting
( which is the order parameter of quantum coherence). Our formulas naturally
give rise to the dilute instanton results for the order parameter both at 
finite and vanishing temperature\cite{bray-moore}. In this sense our attempt 
to reproduce the physical results of the Caldeira-Leggett model without 
referring to an explicit model Lagrangian has been  successful at least in the analyses of quantum coherence, both at $T=0$\cite{fuji} and  $T\neq 0$.

Of course, the formulation of Caldeira and Leggett\cite{caldeira} is flexible 
enough and it is applicable to many other physical contexts. Nevertheless, 
for those who
wonder if one can analyze some physical processes without an infinite number 
of  oscillators, our reformulation of the quantum tunneling with dissipation 
on the basis of fluctuation-dissipation theorem and dispersion relations 
may give an answer by showing such possibility as well as limitations. 

Another purpose  of the present paper is to present  a generalization of 
the fluctuation-dissipation theorem of Callen-Welton to the case of fermionic 
dissipation (or fluctuation). Since fermions are basically quantum 
mechanical, the notion of 
fermionic dissipation is characteristically quantum mechanical. We emphasize 
that a mere excitation of a fermion from one state to another does not imply 
fermionic fluctuation in the present context; the force fields or currents
involved should be fermionic , though the fermionic excitation may be an
effective one or a quasi-particle. In full quantum theory, both of the 
bosonic and fermionic modes can be equally added to or removed from the 
system we are interested in, and a fully quantum mechanical fluctuation-dissipation theorem should be able to handle the fermionic fluctuation and, 
consequently, fermionic dissipation. 

In the context of condensed matter physics, we usually measure {\em bosonic}
quantities such as voltage or electric current even if the elementary 
process involves  the transfer of fermions. In such a case , the conventional 
bosonic fluctuation-dissipation theorem is applicable. It is our hope that 
a properly defined treatment of elementary transfer  processes  may lead 
to an application of the  notion of fermionic fluctuation or dissipation in 
the future. 

As for the  physical phenomena where the fermionic fluctuation-dissipation theorem may have some relevance, we note the fermion emission from a black hole\cite{wald} and the fermion production in an accelerated frame\cite{takagi2}.  The notion of quantum noise plays a fundamental role in the analyses of these processes, and thus the fermionic fluctuation-dissipation theorem may provide a convenient framework to describe some general features of these interesting quantum processes. Another area of physics where fermionic fluctuation may play a
 role is the theory of supersymmetry ( or Boson-Fermion symmetry) \cite{wess}.
 The basic current of supersymmetry is {\em fermionic}, and thus the thermal 
average of a product of such currents ( or related gravitino fields ) inevitably leads to a notion of  fermionic fluctuation. Though we do not know 
a specific application of the fermionic fluctuation-dissipation theorem at this moment, our formulation may turn out to be  useful in the future analyses of 
supersymmetry in a multi-particle thermal setting , where supersymmetry is 
known to be inevitably broken by thermal effects.

\end{document}